\documentclass[12pt]{article}

\usepackage{amsmath,amssymb,color}
\usepackage{pstricks}
\usepackage{graphicx}

\newcommand{\non}{\nonumber}

\newcommand\shalf{\ensuremath{{\scriptstyle\frac{1}{2}}}}

\newcommand{\rem}[1]{}
\newtheorem{theorem}{Theorem}

\DeclareMathAlphabet{\mathbi}{OML}{cmm}{b}{it} 
\newcommand{\bx}{\mathbi{x}}

\newcommand{\bel}{\begin{equation}\label}
\newcommand{\ee}{\end{equation}}
\newcommand{\ben}{\begin{enumerate}}
\newcommand{\een}{\end{enumerate}}
\newcommand{\bde}{\begin{description}}
\newcommand{\ede}{\end{description}}
\newcommand{\bit}{\begin{itemize}}
\newcommand{\eit}{\end{itemize}}
\newcommand{\bc}{\begin{center}}
\newcommand{\ec}{\end{center}}
\newcommand{\bB}{\mbox{\boldmath$B$}}
\newcommand{\bdb}{\mbox{\boldmath$b$}}
\newcommand{\bD}{\mbox{\boldmath$\mathcal{D}$}}

\newcommand{\ba}{\mathbi{a}}

\newcommand{\br}{\mathbi{r}}
\newcommand{\bR}{\mathbi{R}}
\newcommand{\bfR}{\mathfrak{R}}

\newcommand{\bq}{\mathbi{q}}

\newcommand{\bfq}{\mathfrak{q}}
\newcommand{\bhfq}{\hat{\mathfrak{q}}}
\newcommand{\bhfp}{\hat{\mathfrak{p}}}

\newcommand{\bfB}{{\mbox{\boldmath$\mathfrak{B}$}}}

\newcommand{\bfr}{\mathfrak{r}}

\newcommand{\cast}{\circledast}
\newcommand{\bA}{\mbox{\boldmath$\mathcal{A}$}}
\newcommand{\bdB}{\mbox{\boldmath$\mathcal{B}$}}
\newcommand{\bdD}{\mbox{\boldmath$\mathcal{D}$}}
\newcommand{\bdJ}{\mbox{\boldmath$\mathcal{J}$}}

\newcommand{\bhn}{\hat{\mathbi{n}}}
\newcommand{\bu}{\mathbi{u}}

\newcommand{\bU}{\mbox{\boldmath$\mathcal{U}$}}

\newcommand{\bhB}{\mbox{\boldmath$\hat{\mathcal{B}}$}}

\newcommand{\bom}{\mbox{\boldmath$\omega$}}

\newcommand{\bcapom}{\mbox{\boldmath$\Omega$}}

\newcommand{\bchi}{\mbox{\boldmath$\chi$}}
\newcommand{\bhchi}{\boldsymbol{\hat{\chi}}}

\newcommand{\bk}{\mbox{\boldmath$k$}}

\newcommand{\beq}{\begin{eqnarray}\label} 
\newcommand{\eeq}{\end{eqnarray}} 
\newcommand{\bn}{\mathbi{\hat{n}}}
\newcommand{\bnabla}{\mbox{\boldmath$\nabla$}}
\makeatletter
\@addtoreset{equation}{section}

\makeatother

\begin{document}
\sf
\bc
\textbf{\large The gradient of potential vorticity, quaternions\\
and an orthonormal frame for fluid particles}
\par\vspace{7mm}
\textbf{\large J. D. Gibbon and D. D. Holm}
\par\vspace{7mm}
Department of Mathematics, Imperial College London SW7 2AZ, UK\\
{\small email: j.d.gibbon@ic.ac.uk and d.holm@ic.ac.uk}
\ec

\vspace{5mm}

\bc
\textit{Dedicated to Raymond Hide on the occasion of his 80th birthday.}
\ec

\vspace{5mm}

\begin{abstract}
The gradient of potential vorticity (PV) is an important quantity because 
of the way PV (denoted as $q$) tends to accumulate locally in the oceans 
and atmospheres. Recent analysis by the authors has shown that the vector 
quantity $\bdB = \bnabla q\times \bnabla\theta$ for the three-dimensional 
incompressible rotating Euler equations evolves according to the same 
stretching equation as for $\bom$ the vorticity and $\bB$, the magnetic 
field in magnetohydrodynamics (MHD). The $\bdB$-vector therefore acts 
like the vorticity $\bom$ in Euler's equations and the $\bB$-field in MHD. 
For example, it allows various analogies, such as stretching dynamics, 
helicity, superhelicity and cross helicity. In addition, using quaternionic 
analysis, the dynamics of the $\bdB$-vector naturally allow the construction 
of an orthonormal frame attached to fluid particles\,; this is  designated 
as a quaternion frame. The alignment dynamics of this frame are particularly 
relevant to the three-axis rotations that particles undergo as they traverse 
regions of a flow when the PV gradient $\bnabla q$ is large.
\end{abstract}

\vspace{1cm}


\newpage

\section{\large\textsf{Introduction}}\label{intro}

The ideas in this paper weave together two strands of research on the Euler fluid 
equations recently pursued by the authors. Both of these required the use of 
Ertel's Theorem \cite{Er42} -- always a favourite topic with Raymond -- and 
were discussed at length with him during their development. 

The first and latest strand of research concerns the evolution of the gradient 
of potential vorticity ($\bnabla q$) and the gradient of potential temperature 
($\bnabla\theta$) \cite{GH10}. In the case of the Euler equations, while both 
$q$ and $\theta$ are material constants, the evolution of their gradients 
involves the strain and rotation rates of the flow. Physically, understanding 
the behaviour of $\bnabla q$ in the atmosphere and the oceans is of paramount 
importance because potential vorticity tends to accumulate into localised spatial 
regions (patches) with sharp edges, where the magnitude $|\bnabla q|$ is much 
larger than its average value \cite{HMR85}. In this regard, the divergenceless 
vector combination $\bdB = \bnabla q \times\bnabla \theta$ is a natural choice, 
because it leads to an evolution equation identical to that for either the 
vorticity in the incompressible three-dimensional Euler equations, or the 
magnetic field in an electrically conducting fluid. As a consequence, $\bdB$ 
may undergo the same violent stretching and twisting associated with the 
vorticity field in three-dimensional turbulence, or with magnetic field 
lines in magnetohydrodynamics (MHD), particularly if $\bdB$ were to align with 
an eigenvector of the $3\times3$ strain-rate matrix $S$ associated with the 
fluid motion.

The second strand of research involves the use of quaternions in identifying an 
ortho-normal frame attached to fluid particles in an Euler flow and whose dynamics 
represent the tumbling of the particle as it undergoes three-dimensional rotations 
during its flight  \cite{GH06,GHKR}. The vector $\bdB$ turns out to be an ideal 
candidate for the construction of this ortho-normal frame. This would be applicable 
to the dynamics of particles travelling through regions of the oceans or atmospheres 
which have a high values of $|\bnabla q|$.
 
\subsection{\textsf{\large Potential vorticity gradient for the incompressible Euler}}\label{qintro}

Consider the dimensionless form of the Euler equations for incompressible, stratified 
and rotating flow  
\bel{compeul1}
\frac{D\bu}{Dt} + 2\bcapom\times\bu + a_{0}\bk\theta = -\bnabla p\,,\qquad\qquad 
\frac{D~}{Dt} = \frac{\partial~}{\partial t} + \bu\cdot\bnabla\,,
\ee
in which the potential temperature $\theta(\bx,\,t)$ evolves passively according to 
\bel{compeul2}
\frac{D\theta}{Dt} = 0\,.
\ee
In (\ref{compeul1}), the vectors $\bcapom$ and $\bk$ are the rotation rate and vertical 
direction, respectively, and the scalar $a_{0}$ is a dimensionless constant. Information 
about $\bnabla\theta$ would be needed to discuss how $\theta(\bx,\,t)$ might accumulate 
into large local concentrations. This is best studied in the context of potential vorticity 
defined by
\bel{qdef}
q := \bom_{rot}\cdot\bnabla\theta
\ee
where $\bom_{rot}$ is defined as $\bom_{rot} := \bom + 2\bcapom$ and $\bom :={\rm curl}\,\bu$ 
denotes the fluid vorticity. Ertel's theorem says that the material time derivative $D/Dt$ 
and the vector field $\bom_{rot}\cdot\bnabla$ operating on a scalar function commute with 
each other under an Euler flow \cite{Er42}
\bel{q1A}
\frac{Dq}{Dt} = 
\left(\frac{D\,\bom_{rot}}{Dt} 
- \bom_{rot}\cdot\bnabla\bu \right)\cdot\bnabla \theta
+ \bom_{rot}\cdot\bnabla\left(\frac{D\theta}{Dt}\right)\,.
\ee
When $\bom_{rot}$ obeys the incompressible Euler equations $(\bnabla^{\perp} = 
\left(\partial_{y},\,-\partial_{x},\,0\right))$
\bel{q1B}
\frac{D\,\bom_{rot}}{Dt} = \bom_{rot}\cdot\bnabla\bu - a_{0}\bnabla^{\perp}\theta 
\ee
then $q$ is a also material constant because
\bel{q1C}
\frac{Dq}{Dt} = 0\,.
\ee

\par\vspace{-30mm}
\bc
\begin{minipage}[htb]{11cm}
\setlength{\unitlength}{.7cm}
\begin{picture}(11,11)(0,0)
%
\qbezier[400](.5,5)(5,3)(8,3)
\qbezier[400](.5,5)(.75,5.5)(.5,6)
\qbezier[400](.5,6)(2,5.5)(3,5)
\qbezier[400](3,5)(6,5.8)(9,6.5)
\qbezier[400](8,3)(7,3.5)(7,4)
\qbezier[400](7,4)(6.5,4.1)(5.3,4.4)
\qbezier[400](1,3)(3,4)(8,5)
\qbezier[400](1,3)(1,3.5)(.5,4)
\qbezier[400](.5,4)(1,4.2)(1.6,4.5)
\qbezier[200](9,6.5)(8,5.8)(8,5)
\thicklines
\qbezier[400](3.45,3.9)(3.2,4.5)(3,5)
\put(3.75,4.2){\vector(-1,2){.4}}
\put(4.1,4.4){\makebox(0,0)[b]{\scriptsize$\bdB$}}
\put(8.4,3.5){\makebox(0,0)[b]{\scriptsize$\theta = \mbox{const}$}}
\put(-0.2,3.2){\makebox(0,0)[b]{\scriptsize$q = \mbox{const}$}}
\put(5.7,3.4){\makebox(0,0)[b]{\scriptsize$\bnabla\theta\!\nearrow$}}
\put(7,5.1){\makebox(0,0)[b]{\scriptsize$\nwarrow\!\!\bnabla q$}}
\end{picture}
\end{minipage}
\ec
\par\vspace{-15mm}\noindent
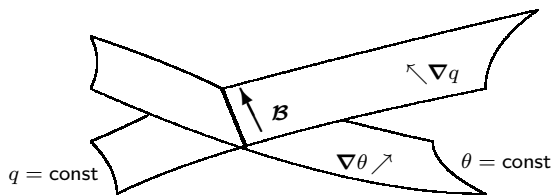
{\textbf{Figure 1\,:} \textit{\small The vector $\bdB = \bnabla q\times\bnabla\theta$ 
is tangent to the curve defined by the intersection of the two surfaces $q = \mbox{const}$ 
and $\theta = \mbox{const}$ in three dimensions.}\label{fig1}}
\par\vspace{5mm}\noindent
To achieve this, the following divergence-free flux vector is constructed \cite{GH10}
\bel{Bdef}
\bdB = \bnabla Q(q)\times\bnabla \theta \,,
\ee
where $Q(q)$ is any smooth function of $q$. The vector $\bdB$, as in Figure 1, lies along 
the intersection of iso-surfaces of $q$ and $\theta$ and could be thought of as pointing 
along the tangent to an iso-PV curve on a level set of temperature, or vice versa. Thus, 
the PV function $Q(q)$ is the stream function for the flux vector $(\bdB)$ on a level set 
of potential temperature $(\theta)$. This observation results in the remarkably simple 
evolution equation
\beq{stretch1}
\frac{\partial\bdB}{\partial t} - \mbox{curl}\,(\bu\times\bdB) = 
-\bnabla(qQ'\mbox{div}\,\bu)\times\bnabla\theta\,.
\eeq
The cross-product combination is special\,: an appendix in \cite{GH10} contains a proof 
of (\ref{stretch1}) performed using differential geometry and also by conventional vector 
analysis. It follows from (\ref{stretch1}) that $\bdB$ satisfies the stretching relation
\bel{sreuler}
\frac{D\bdB}{Dt} = \bdB\cdot\bnabla\bu  - \bdB\,\mbox{div}\,\bu - \bnabla(qQ'\mbox{div}\,\bu)\times\bnabla\theta\,,
\ee
whose properties will be discussed in the next section\footnote{\sf In \cite{GH10} the original references
have been discussed (\cite{KurgTat87,KurgPis00,Kurg02}) in which (\ref{stretch1}) had been derived for the 
incompressible Euler equations where the right hand side turned out to be zero because $\mbox{div}\,\bu = 0$.
In \cite{GH10} the argument is extended to both the Navier-Stokes and hydrostatic viscous primitive equations.}.

\subsection{\textsf{Stretching, helicity, superhelicity and cross helicity}}\label{super}

In the incompressible case where $\mbox{div}\,\bu = 0$, equation (\ref{sreuler}) 
simplifies so that the divergenceless vector $\bdB$ satisfies the same stretching 
equation as that for the vorticity $\bom$\,; namely 
\bel{Bstretch}
\frac{D\bdB}{Dt} = \bdB\cdot\bnabla\bu
\qquad\hbox{with}\qquad
\mbox{div}\, \bdB = 0\,.
\ee
This leads immediately to
\bel{ev1}
\frac{D~}{Dt}|\bdB|^{2} = \bdB\cdot S\bdB = \lambda^{\tiny (S)}|\bdB|^{2}\,,
\ee
where $\lambda^{\tiny(S)}(\bx,\,t)$ is an estimate for an eigenvalue of the rate of strain 
matrix $S$ and lies within its spectrum.  Alignment of $\bdB$ with a positive (negative) 
eigenvector of $S$ may produce violent growth (decay) thus reproducing the stretching 
mechanism that produce the very large vorticity intensities that can develop locally in 
the early and intermediate stages of turbulence. 
  
Moffatt's analogy between vorticity and magnetic field \cite{HKM1}, and his detailed 
discussion of the topology of magnetic field lines, is based on the concept of helicity 
which for us requires the existence of a vector potential $\bA$ defined by
\bel{Adef}
\bA = \shalf\big(Q\bnabla\theta - \theta\bnabla Q\big) + \bnabla\psi\,,
\ee
in terms of the dynamical quantities $Q$, $\theta$ and a potential $\psi$. Then the 
helicity $H$, defined by
\bel{heldef}
H = \int_{V}\bA\cdot\bdB\,dV 
= \oint_{\partial V} \psi \bdB\cdot\bn\,dS\,,
\ee
measures the number of linkages of the field lines of $\bdB$ with themselves.
The time derivative of helicity under the flow of the Euler equations is given by
\bel{hel-dot}
\frac{dH}{dt}  
= \oint_{\partial V} \bigg[-(\bA\cdot\bdB)\, \bu\cdot\bn
+ \Big(\frac{D \psi}{Dt}\Big)\,  \bdB\cdot\bn\bigg]\,dS\,,
\ee
which would vanish for either homogeneous or periodic boundary conditions. 
For the Euler equations, $\bu\cdot\bn = 0$ is imposed on a fixed boundary. 
However, $H\neq 0$ and $dH/dt \ne 0$ would be possible for topography in 
which $\bdB\cdot\bn \ne 0$, so the boundaries play the only role in 
allowing linkages in the $\bdB$-field as there is no other source of 
helicity. 

Hide's intriguing concept of a \textit{super-helicity}, which measures the 
linkages of the field lines of $\bdJ:=\mbox{curl}\,\bdB$ with itself, 
may be introduced for the $\bdJ$-vector as in for MHD \cite{Hide1989}. 
Super-helicity is defined as
\bel{superhel}
\mathcal{S}= \int_{V}\bdB\cdot\bdJ\,dV\,.
\ee
After a short computation, the super-helicity dynamics for the $\bdJ$-vector 
comes out to be
\bel{superheldyn}
\frac{d\mathcal{S}}{dt}
= \int_{V} 2\bdB\cdot \mbox{curl}^2(\bu\times\bdB)\,dV
+ \oint_{\partial V} \Big[(\bu\cdot\bdJ)\,  \bdB\cdot\bn 
- (\bdJ\cdot\bdB)\, \bu\cdot\bn\Big]\,dS\,,
\ee
which, unlike the helicity $H$, has both volume and surface sources. Likewise the 
\textit{cross helicity} for the $\bdB$-vector can be introduced
\bel{crosshel-def}
\mathcal{C} = \int_{V}\bu\cdot\bdB\,dV 
\ee
in analogy with the corresponding quantity in MHD.  Another short computation produces 
the dynamics of the cross helicity ($\mathbf{R}$ is the vector potential such that 
$\mbox{curl}\,\mathbf{R} = 2\bcapom$) 
\bel{cross-heldyn}
\frac{d\,\mathcal{C}}{dt}
= -a_{0}\int_{V} ( \theta\,\bdB\cdot\bk) \,dV + \oint_{\partial V}
\Big(-p + \bu\cdot\mathbf{R} + \shalf u^{2}\Big)\,\bdB\cdot\bn \,dS\,,
\ee
which again has both volume and surface sources. 

\rem{Finally it may be tempting to extend the parallel between $\bom$ and $\bdB$ further by 
defining the sequence of scalar functions $q_{n} = \bdB_{n-1}\cdot\bnabla\theta$ with 
$\bdB_{0} = \bom$, and then constructing an infinite set $\bdB_{n} = \bnabla Q_{n}(q_{n})
\times\bnabla\theta$ which all obey the same stretching relation (\ref{sreuler}). However, 
this construction, while elegant, contains no new information and is equivalent to computing 
the higher terms in a Taylor series expansion of an arbitrary function of $q$ in the initial 
conditions.}

\subsection{\textsf{Quaternions and an attached orthonormal frame}}\label{quatintro}

\par\vspace{-6mm}\noindent
\bc
\begin{minipage}[c]{.75\textwidth}
\begin{pspicture}
\psframe(0,0)(5,5)
\thicklines
\qbezier(0,1)(4,2.5)(0,4)
\thinlines
\put(2.01,2.43){\makebox(0,0)[b]{$\bullet$}}
\put(1.4,2.5){\makebox(0,0)[b]{\scriptsize$(\bx_{1},t_{1})$}}
\thinlines
\put(2,2.5){\vector(0,1){1}}
\put(2,3.7){\makebox(0,0)[b]{$\bhB$}}
\put(2,2.5){\vector(-2,-1){1}}
\put(.7,1.8){\makebox(0,0)[b]{$\bhchi_{a}$}}
\put(2,2.5){\vector(1,0){1}}
\put(3.8,2.4){\makebox(0,0)[b]{$\bhB\times\bhchi_{a}$}}
\thicklines
\qbezier(7,1)(6,2.5)(8,4)
\thinlines
\put(6.77,2.45){\makebox(0,0)[b]{$\bullet$}}
\put(6.2,2.5){\makebox(0,0)[b]{\scriptsize$(\bx_{2},t_{2})$}}
\thicklines
\qbezier[50](2,2.5)(3,.5)(6.7,2.5)
\thinlines
\put(6.73,2.5){\vector(1,4){.3}}
\put(7,3.8){\makebox(0,0)[b]{$\bhB$}}
\put(6.7,2.5){\vector(4,-1){1}}
\put(8.4,2.8){\makebox(0,0)[b]{$\bhB\times\bhchi_{a}$}}
\put(6.7,2.5){\vector(4,1){1}}
\put(8.1,2.1){\makebox(0,0)[b]{$\bhchi_{a}$}}
\put(3,1.2){\vector(1,0){.6}}
\put(4,.7){\makebox(0,0)[b]{\scriptsize trajectory of tracer particle}}
\put(4.5,1.4){\vector(4,1){.6}}
\thinlines
\end{pspicture}
\end{minipage}
\ec
\bc
\vspace{-3mm}
\begin{minipage}[r]{\textwidth}
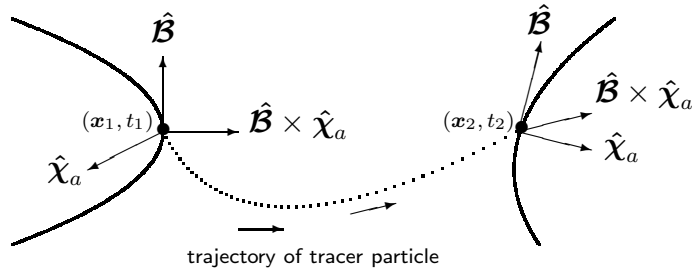
\textbf{Figure 2\,:} \textit{\small  The solid curves represent characteristic curves 
$\bhB = d\bx/ds$ ($s$ is arc length) to which $\bhB$ is a unit tangent vector. 
The quaternion-frame orientation $(\bhB,\,\bhchi_{a},~\bhB\times\bhchi_{a})$ 
is shown at the two space-time points $(\bx_{1},t_{1})$ to $(\bx_{2},t_{2})$; 
note that this is not the Frenet-frame corresponding to the particle path but 
to curves $\bhB = d\bx/ds$. The dotted line represents the tracer particle 
$(\bullet)$ path.} 
\end{minipage}
\ec
More than one hundred and fifty years ago William Rowan Hamilton invented quaternions 
as a means of representing a composition of rotations. For most of this period they 
have been under-appreciated, yet they have recently undergone a spectacular renaissance 
due to their efficacy in certain applications in avoiding the difficulties incurred 
at the north and south poles\footnote{\sf Computations with Euler angles often suffer 
from ``gyro-lock'' because of singularities at the poles of the spherical angular 
coordinate system where the azimuthal angle is undefined.}  when Euler angles are used 
in computing the dynamics of objects undergoing three-axis rotations \cite{Ha06}. In 
particular, quaternions now lie at the heart of many modern inertial guidance systems 
where tracking the paths and the orientation of aircraft and satellites is of importance 
\cite{Ku99}. They are also used in the graphics community to control the orientation of 
tumbling objects in computer animations \cite{Ha06}. 

A natural question is whether quaternions are useful in tracking the angular velocity and 
orientation of Lagrangian particles in fluid dynamics. Experiments in turbulent flows 
have now reached the stage where the trajectories of tracer particles can be detected 
at high Reynolds numbers \cite{La01,Vo02,RMCB05,Eck06}. Numerical differentiation of 
these trajectories gives information about the Lagrangian velocity and acceleration 
of the particles and also the curvature of the particle paths \cite{Eck06}. 

Conventional practice has been to consider the Frenet-frame of a trajectory. This consists 
of a unit tangent vector, a normal and a bi-normal, which are used to represent the pitch, 
yaw and roll of the motion. While the Frenet-frame describes the path, it ignores the 
rotational dynamics of the particle. To account for this, another ortho-normal 
frame associated with the motion of a Lagrangian fluid particle -- designated the 
\textit{quaternion-frame} -- has been introduced by the authors \cite{GH06}. This frame 
moves with the particles, but its evolution derives from the fluid equations of motion. 
In the context of the incompressible Euler equations, the idea of the quaternion frame 
depends on the existence of vectors for which there exists a Lagrangian equation of 
motion\footnote{\sf The eigenvectors of the rate of strain matrix $S$ are excluded 
for this reason.}. The natural triplet is $\{\bom,\,\bhchi,\,\bom\times\bhchi\}$ 
where $\bchi = \bom\times S\bom$ -- see \cite{GH06,GHKR}. For the incompressible Euler 
equations the natural candidate for the $\bdB$-stretching equation (\ref{Bstretch}) is 
the triplet 
\bel{triplet1}
\{\bhB,\,\bhchi_{a},\,\bhB\times\bhchi_{a}\}\qquad\mbox{where}\qquad
\bchi_{a} = \bhB\times \ba
\ee
where the $\ba$-label has its origins in the definition $\ba := \bdB\cdot\bnabla\bu$. 
The Lagrangian 
equations of motion for this triplet is derivable through a quaternionic formulation of 
the Lagrangian equation of motion for $\bdB$ given in (\ref{Bstretch}).

\section{\large\textsf{Quaternions, rigid body rotations and their properties}}\label{quatprop}

The material in this section provides the reader with a definition of quaternions, 
together with a pr\`ecis of their multiplication rules and properties. 

The literature on rotations in rigid body mechanics is replete with explicit formulae 
relating the Euler angles and what are called the Cayley-Klein parameters of a 
rotation \cite{Whitt44}. The complicated inter-relations that are unavoidable when Euler 
angle formulae are used can be avoided when quaternions are used \cite{Klein04,Ta1890}\,: 
for a more modern context see Holm \cite{Holmgm1,Holmgm2} and Marsden and Ratiu \cite{MaRa1994}. 
\par\smallskip
In terms of any scalar $p$ and any 3-vector $\bq$, the 4-vector quaternion $\bfq = [p,\,\bq]$ 
is defined as (Gothic fonts denote quaternions)
\bel{quatdef1}
\bfq = [p,\,\bq] = p{\sf I} - \sum _{i=1}^{3}q_{i}\sigma_{i}\,,
\ee
where $\{\sigma_{1},\,\sigma_{2},\,\sigma_{3}\}$ are the three Pauli spin-matrices defined by
\bel{psm1}
\sigma_{1} = \left(\begin{array}{rr}
0 & i\\
i & 0
\end{array}\right)\,,
\hspace{1cm}
\sigma_{2} = \left(\begin{array}{rr}
0 & 1\\
-1 & 0
\end{array}\right)\,,
\hspace{1cm}
\sigma_{3} = \left(\begin{array}{rr}
i & 0\\
0 & -i
\end{array}\right)\,,
\ee
and ${\sf I}$ is the $2\times 2$ unit matrix. A rule (denoted as $\cast$) 
for multiplication of quaternions can be derived from the multiplication 
rule for the Pauli matrices $\sigma_{i}\sigma_{j} =  - 
\delta_{ij}{\sf I}-\epsilon_{ijk}\sigma_{k}$, as
\bel{quatdef2}
\bfq_{1}\cast\bfq_{2} = [p_{1}p_{2} - \bq_{1}\cdot\bq_{2},\,p_{1}\bq_{2} + p_{2}\bq_{1} + 
\bq_{1}\times\bq_{2}]\,.
\ee
Thus, the multiplication of quaternions is associative, but it is not commutative.

Let $\bhfp = [p,\,\bq]$ be a unit quaternion, satisfying $p^2 + q^2 = 1$. Its 
inverse $\bhfp^{*} = [p,\,-\bq]$ satisfies $\bhfp \cast\bhfp^{*} = [p^2 + q^2,\,0] 
= [1,0]$. A pure quaternion has a zero scalar entry, such as $\bfr = [0,\,\br]$. 
Hamilton called his pure quaternions \emph{vectors}. Pure quaternions transform 
among themselves as $\bfr = [0,\,\br] \to \bfR = [0,\,\bR]$ under
\bel{r1}
\bfR = \bhfp\cast\bfr\cast\bhfp^{*}\,.
\ee
This associative product can be written as
\bel{r2}
\bfR = \bhfp\cast\bfr\cast\bhfp^{*} 
= [0,\, (p^{2}-q^{2})\br + 2p(\bq\times\br)+ 2\bq(\br\cdot\bq)]\,.
\ee
Choosing $p=\pm\cos\shalf\theta$ and $\bq = \pm\,\bhn\sin\shalf\theta$, where $\bhn$ 
is the unit normal to $\br$, we find that 
\bel{r3}
\bfR = \bhfp\cast\bfr\cast\bhfp^{*} = [0,\, \br\cos\theta + (\bhn\times\br) \sin\theta] 
\equiv O(\theta\,,\bhn)\br\,,
\ee
where
\bel{r4}
\bhfp = \pm [\cos\shalf\theta,\,\bhn\sin\shalf\theta]\,.
\ee
Equation (\ref{r3}) is the Euler-Rodrigues formula for the rotation $O(\theta\,,\bhn)$ 
by an angle $\theta$ of the 3-vector $\br$ about its normal $\bhn$ and the quantities 
$\theta\,,\bhn$ are the Euler parameters. The elements of the unit quaternion $\bhfp$ 
are the Cayley-Klein parameters which are related to the Euler angles \cite{Whitt44}. 
When $\bhfp$ is time-dependent, the Euler-Rodrigues formula in (\ref{r3}) is
\bel{rot2}
\bfR(t) = \bhfp\cast\bfr\cast\bhfp^{*}
\hspace{1cm}\Rightarrow\hspace{1cm}
\bfr = \bhfp^{*}\cast\bfR(t)\cast\bhfp\,.
\ee
It is necessary to use the property of the pure quaternion $\bfR^{*} = -\bfR$ 
to obtain the time derivative of $\dot{\bfR}$
\beq{rot3}
\dot{\bfR}(t) 
&=& (\dot{\bhfp}\cast\bhfp^{*})\cast\bfR - ((\dot{\bhfp}\cast\bhfp^{*})\cast\bfR)^{*}\,.
\eeq
The quaternion $\bhfp = [p,\,\bq]$ is of unit length and so $p\dot{p} + q\dot{q} = 0$, which 
means that $\dot{\bhfp}\cast\bhfp^{*}$ is also a pure quaternion 
\bel{rot4}
\dot{\bhfp}\cast\bhfp^{*} = [0,\,\shalf\bcapom_{0}(t)]\,.
\ee
The 3-vector entry in (\ref{rot4}) defines the angular frequency $\bcapom_{0}(t)$ as 
$\bcapom_{0} = 2(-\dot{p}\bq +\dot{\bq}p - \dot{\bq}\times\bq)$ thereby giving the 
well-known formula for the rotation of a rigid body
\bel{rot5}
\dot{\bR} = \bcapom_{0}\times\bR\,.
\ee 
For a Lagrangian particle, the equivalent of $\bcapom_{0}$ is the Darboux vector $\bD_{a}$ 
in Theorem \ref{abthm} in the next section.

\section{\large\textsf{An ortho-normal frame and particle trajectories}}\label{ortho-normal}

Having set the scene in \S\ref{quatprop} by describing some of the essential properties 
of quaternions, it is now time to apply them to the Lagrangian relation (\ref{Bstretch}) 
between the two vectors $\bdB$ and $\ba$ 
\bel{Bequn}
\frac{D\bdB}{Dt} = \ba : = \bdB\cdot\bnabla\bu\,.
\ee
It will turn out below that a knowledge of $D\ba/Dt$ is needed. Ertel's Theorem is applicable 
and the result becomes a version of Ohkitani's relation \cite{Ohk93}
\beq{ertelex1}
\frac{D~}{Dt}\big(\bdB\cdot\bnabla\bu\big) &=& \bdB\cdot\bnabla
\left(\frac{D\bu}{Dt}\right)\\
&=& - P\bdB - \bdB\cdot\bnabla\left(2\bcapom\times\bu + a_{0}\bk\theta\right)
\eeq
where the Hessian matrix of the pressure $p$ is defined as
\bel{Hessdef}
P = \frac{\partial^{2}p}{\partial x_{i}\partial x_{j}}\,.
\ee
Thus we can define a vector $\bdb$ such that
\bel{ertelex2}
\frac{D\ba}{Dt} = \bdb := - P\bdB - 
\bdB\cdot\bnabla\left(2\bcapom\times\bu + a_{0}\bk\theta\right)\,.
\ee
Through the  multiplication rule in (\ref{quatdef2}) quaternions appear in the 
decomposition of the 3-vector $\ba$ into parts parallel and perpendicular to another 
vector, which we choose to be $\bdB$. This decomposition is expressed as
\bel{decom1}
\ba = \alpha_{a}\bdB + \bchi_{a}\times\bdB = [\alpha_{a},\,\bchi_{a}]\cast[0,\,\bdB]\,,
\ee
where the scalar $\alpha_{a}$ and 3-vector $\bchi_{a}$ are defined as
\bel{la2a}
\alpha_{a} = \mathcal{B}^{-1}(\bhB\cdot\ba)\,,\hspace{2cm}
\bchi_{a} = \mathcal{B}^{-1}(\bhB\times\ba)\,.
\ee
Equation (\ref{decom1}) thus shows that the quaternionic product is summoned in 
naturally.\footnote{\sf With reference to \S\ref{quatprop}, the Cayley-Klein parameters 
of the quaternion $\bfq = [\alpha,\,\bchi]$ are
$$
\bhfq = \left[\frac{\alpha}{\alpha^2 + \chi^2},\, \frac{\bchi}{\alpha^2 + \chi^2}\right]\,.
$$} 
It is now easily seen that $\alpha_{a}$ is the \emph{growth rate} of the scalar magnitude 
($\mathcal{B} =|\bdB|$) which obeys
\bel{la3}
\frac{D\mathcal{B}}{Dt} = \alpha_{a}\mathcal{B}\,,
\ee
while $\bchi_{a}$, the \emph{swing rate} of the unit tangent vector $\bhB = \bdB \mathcal{B}^{-1}$, 
satisfies 
\bel{la5}
\frac{D\bhB}{Dt} = \bchi_{a}\times \bhB\,.
\ee
Now define the two quaternions
\bel{ls6}
\bfq_{a} 
= [\alpha_{a},\,\bchi_{a}]\,,
\hspace{2cm}
\bfB = [0,\,\bdB]\,,
\ee
so (\ref{Bequn}) can automatically be re-written in the quaternion form
\bel{lem1}
\frac{D\bfB}{Dt} 
= \bfq_{a}\cast\bfB\,.
\ee
Moreover, because of (\ref{ertelex2}), exactly as for $\bfq_{a}$, a quaternion $\bfq_{b}$ 
can be defined which is based on the variables 
\bel{alphachibdef}
\alpha_{b} = \mathcal{B}^{-1}(\bhB\cdot\bdb)\,,
\hspace{3cm}
\bchi_{b} = \mathcal{B}^{-1}(\bhB\times\bdb)\,,
\ee
where
\bel{qbdef}
\bfq_{b} = [\alpha_{b},\,\bchi_{b}]\,.
\ee
The 3-vector $\bdb = D\ba/Dt$ admits a decomposition similar to that for $\ba$ as in (\ref{decom1}) 
\bel{la9}
\frac{D^{2}\bfB}{Dt^{2}} 
= [0,\,\bdb] 
= [0,\, \alpha_{b}\bdB + \bchi_{b}\times\bdB ] 
= \bfq_{b}\cast\bfB \,.
\ee
Using the associativity property, compatibility of (\ref{la9}) 
and (\ref{lem1}) implies that ($\mathcal{B} = |\bdB| \neq 0$)
\bel{la10}
\left(\frac{D\bfq_{a}}{Dt} + \bfq_{a}\cast\bfq_{a} 
-\bfq_{b}\right)\cast\bfB = 0\,.
\ee
This establishes a Riccati relation between $\bfq_{a}$ and $\bfq_{b}$
\bel{Ric1}
\frac{D\bfq_{a}}{Dt} + \bfq_{a}\cast\bfq_{a} = \bfq_{b}
\,,
\ee
with components
\bel{Ric1-comp}
\frac{D}{Dt} [\alpha_{a},\,\bchi_{a}]
+ [\alpha_{a}^2-\chi_{a}^2,\,2\alpha_{a}\bchi_{a}] 
= [\alpha_{b}\,,\chi_{b}]
\,,
\ee
where $\chi_{a} = |\bchi_{a}|$. There follows a Theorem on a Lagrangian particle 
undergoing fluid motion that is equivalent to the well-known formula (\ref{rot5}) 
for a rigid body undergoing rotation about its center of mass\,:
\begin{theorem}[Alignment dynamics]
\label{abthm}
The ortho-normal quaternion-frame $\big(\bhB,\,\bhchi_{a},\,\bhB\times\bhchi_{a}\big)$ 
has Lagrangian time derivatives 
\beq{abframe3}
\frac{D\bhB}{Dt}&=& \bD_{a}\times\bhB\,,\\
\frac{D(\bhB\times\bhchi_{a})}{Dt} &=& \bD_{a}\times(\bhB\times\bhchi_{a})\,,\label{abframe4}
\\
\frac{D\bhchi_{a}}{Dt} &=& \bD_{a}\times\bhchi_{a}\,,\label{abframe5}
\eeq
where the Darboux angular velocity vector $\bD_{a}$ is defined as
\bel{abframe6}
\bD_{a} = \bchi_{a} + \frac{c_{b}}{\chi_{a}}\bhB\,,\hspace{2cm}
c_{b} = \bhB\cdot(\bhchi_{a}\times\bchi_{b})\,,
\ee
and the quantities $[\alpha_{a},\bchi_{a}]$ and $[\alpha_{b},\bchi_{b}]$ are defined in 
(\ref{la2a}) and (\ref{alphachibdef}).
\end{theorem}
\par\smallskip\noindent
\textbf{Remark\,:} The Darboux vector $\bD_{a}$ is driven by the 3-vector $\bdb=D\ba/Dt$ which sits 
in $c_{b}$ in (\ref{abframe6}). The analogy with rigid body rotation expressed in (\ref{rot5}) 
is clear.
\par\smallskip\noindent
\textbf{Proof\,:} To find an expression for the Lagrangian time derivatives of the components of 
the frame  $(\bhB,\,\bhchi_{a},\,\bhB\times\bhchi_{a})$ requires the derivative of $\bhchi_{a}$. 
To find this, it is necessary to use the fact that the 3-vector $\bdb$ can be expressed in this 
ortho-normal frame as the linear combination
\bel{b1}
\mathcal{B}^{-1}\bdb = \alpha_{b}\,\bhB + c_{\,b}\bhchi_{a} + d_{\,b}(\bhB\times\bhchi_{a})\,.
\ee
where $c_{\,b}$ is defined in (\ref{abframe6}) and $d_{\,b} = -\,(\bhchi_{a}\cdot\bchi_{b})$.
The 3-vector product $\bchi_{b} = \mathcal{B}^{-1}(\bhB\times\bdb)$ yields 
\bel{bchibdef}
\bchi_{b} = c_{\,b}\,(\bhB\times\bhchi_{a}) - d_{\,b}\bhchi_{a}\,.
\ee
When split into components, equation (\ref{Ric1-comp}) becomes
\bel{abframe0}
\frac{D\alpha_{a}}{Dt} = \chi_{a}^{2} - \alpha_{a}^{2} + \alpha_{b}\,
\ee
and
\bel{abframe1}
\frac{D\bchi_{a}}{Dt} = - 2\alpha_{a}\bchi_{a} + \bchi_{b}\,.
\ee
A little more working gives the \emph{alignment dynamics} in 
equations (\ref{abframe3})-(\ref{abframe6}).
\hspace{10mm}$\blacksquare$

\section{\large\textsf{Conclusions}}\label{concl}

When the quaternion approach to rotations outlined in \S\ref{quatprop} is 
applied to the Euler equations it demonstrates that quaternions are a 
natural way of calculating the orientation of Lagrangian particles in 
motion through the concept of ortho-normal quaternion-frames attached 
to each particle. In this particular context, where the $\bdB$-field 
is the vector that helps us understand the evolution of $\bnabla q$ and 
$\bnabla\theta$, knowledge of the quartet of 3-vectors $(\bu,\,\bdB,\,\ba,\,\bdb)$ 
is sufficient for the application of Theorem \ref{abthm}. The complexity 
of the $3D$ Euler equations comes through the ortho-normal dynamics via 
the pressure field. In the present state of knowledge, the projection 
$P\bdB$ that is part of $\bdb$ would have to be found by computational means.

A natural question is whether these ideas can be applied to the Navier-Stokes 
equations? It turns out that the evolution equation for $\bdB$ in the 
incompressible case is the same as that in (\ref{sreuler}) with $\bu$ replaced
by $\bU$ \cite{GH10}.  $\bU$ is a new transport velocity calculated using the 
method of Haynes and McIntyre \cite{HMc90}
\bel{HM1}
q(\bU - \bu) = - \left\{\big[ Re^{-1}\Delta\bu + 2\bcapom\times\bu + 
a_{0}\bk\theta\big]\times\bnabla\theta + 
(\sigma Re)^{-1}\bom\Delta\theta\right\}
\ee
The quaternion procedure can only be pursued to a certain point for the Navier-Stokes 
equations, after which  a difficulty  appears when $D\bU/Dt$ is needed, and we have 
no proper knowledge of this. This is consistent with the objections that $\bU$ is not 
a genuine \textit{physical} velocity but merely a mathematical construction 
\cite{Dan90,Viudez99}. For the Euler equations we know that this is the point 
where the Hessian matrix of the pressure is introduced in equation (\ref{ertelex1}). 
Moreover, the stretching relation (\ref{sreuler}) in this case becomes
\bel{sreuler-Bee}
\frac{D\bdB}{Dt} = \bdB\cdot\bnabla\bU  - \bdB\,\mbox{div}\,\bU
- \bnabla(qQ'\mbox{div}\,\bU)\times\bnabla\theta\,,
\ee
and $\mbox{div}\,\bU\ne0$, which allows richer and potentially more singular alignment 
dynamics than those for the incompressible case discussed here in equations 
(\ref{abframe3})-(\ref{abframe6}). As well as the Navier-Stokes alluded to above, the 
case of the hydrostatic primitive equations have been discussed in this context in \cite{GH10}.

\par\vspace{-2mm}

\bibliographystyle{unsrt}

\end{document}